\documentclass[conference]{IEEEtran}
\IEEEoverridecommandlockouts
\usepackage{tabularx}
\usepackage{booktabs} 
\usepackage{cite}
\usepackage{amsmath,amssymb,amsfonts}
\usepackage{algorithmic}
\usepackage{algorithm}
\usepackage{graphicx}
\usepackage{textcomp}
\usepackage{xcolor}
\usepackage{multirow}
\usepackage{url}
\usepackage{graphicx}
\usepackage{subcaption}
\usepackage{amssymb}

\def\BibTeX{{\rm B\kern-.05em{\sc i\kern-.025em b}\kern-.08em
    T\kern-.1667em\lower.7ex\hbox{E}\kern-.125emX}}
\begin{document}



\title{NavVI: A Telerobotic Simulation with Multimodal Feedback for Visually Impaired Navigation in Warehouse Environments}

\author{
Maisha Maimuna\textsuperscript{*}, Minhaz Bin Farukee\textsuperscript{*}, Sama Nikanfar, Mahfuza Siddiqua, Ayon Roy, Fillia Makedon\\
Department of Computer Science and Engineering\\
The University of Texas at Arlington\\
Arlington, TX, USA\\
\{mxm0641, mxf4424, sxn8789, mxs9653, axr7225\}@mavs.uta.edu, makedon@uta.edu\\
}

\maketitle
\begingroup
\renewcommand\thefootnote{*}
\footnotetext{Both authors contributed equally to this research.}
\endgroup

\begin{abstract} 
Industrial warehouses are congested with moving forklifts, shelves and personnel, making robot teleoperation particularly risky and demanding for blind and low-vision (BLV) operators. Although accessible teleoperation plays a key role in inclusive workforce participation, systematic research on its use in industrial environments is limited, and few existing studies barely address multimodal guidance designed for BLV users. We present a novel multimodal guidance simulator that enables BLV users to control a mobile robot through a high-fidelity warehouse environment while simultaneously receiving synchronized visual, auditory, and haptic feedback. The system combines a navigation mesh with regular re-planning so routes remain accurate avoiding collisions as forklifts and human avatars move around the warehouse. Users with low vision are guided with a visible path line towards destination; navigational voice cues with clockwise directions announce upcoming turns, and finally proximity-based haptic feedback notifies the users of static and moving obstacles in the path. This real-time, closed-loop system offers a repeatable testbed and algorithmic reference for accessible teleoperation research. The simulator’s design principles can be easily adapted to real robots due to the alignment of its navigation, speech, and haptic modules with commercial hardware, supporting rapid feasibility studies and deployment of inclusive telerobotic tools in actual warehouses.

\end{abstract}


\section{Introduction} 
Despite the increasing importance of workplace inclusion, visually impaired (VI) individuals remain underrepresented in industrial warehouses \cite{chhabra2021turning}. The environment in warehouses is dynamic, obstacle-rich and require precise navigation and acute spatial awareness. Most existing works focus on fully autonomous robots that assist with tasks, rather than allowing blind users to directly control or navigate these spaces by themselves \cite{gnecco2023increasing}. Wearable technologies and smart canes provide assistance, but they are not suitable for complex environments where safety, responsiveness, and environmental awareness are crucial \cite{mai2023laser}. This disparity emphasizes the requirement for accessible telerobotic devices that can enable BLV users to conduct spatial tasks autonomously and enhance their chances to engage in the future logistic workforce.\\
Telerobotic systems provide an effective way to enhance accessibility for VI individuals, allowing them to remotely explore and maneuver areas through robotic control \cite{zhang2022telepresence}. However, the majority of current telerobotic interfaces mainly depend on visual cues, thus limiting accessibility for BLV users \cite{livatino2021intuitive}. To address these gaps, we introduce NavVI (Navigation Assistant for Vision-Impaired Users), a simulation-based telerobotic interface designed for warehouse navigation leveraging multimodal feedback. NavVI, developed in Unity, enables users to control a mobile robot using a Sony DualSense controller, providing real-time haptic feedback based on obstacle proximity, audio feedback and clock-based navigational cues, and additional visual aids. Though we focused on developing a system that facilitates efficient navigation without dependency on vision, we integrated high-contrast visual indicators to provide additional assistance to low to moderate vision users \cite{perkinshighcontrast}. While in play mode, the boxes in the shelves are separated using bright yellow color, the goal is represented as bright red and there is a vivid purple path while navigating within the warehouse. Obstacles are categorized into left, center, or right zones according to the robot's relative position, while vibration intensity is modulated through a logarithmic decay function dependent on obstacle proximity.
We strategically utilized simulation in this work to support controlled, repeatable, safe early-stage design and testing. By reducing the risks connected with physical robots, we are facilitating (1) safety in developing prototypes by means of a consistent environment and feedback systems,  (2) control and repeatability by means of a warehouse-scale configuration that is difficult to replicate in laboratory environments, and (3) a complete testbed for future user studies. The contributions of our work are as follows. 

\begin{itemize}
    \item We present NavVI, a simulation-based telerobotic system that facilitates warehouse navigation for visually impaired users with the help of multimodal feedback.
    \item A direction and proximity aware haptic feedback system that uses logarithmic function to map obstacle distance.
    \item A scalable simulation framework incorporating NavMesh-based path planning to facilitate future assessment and real-world adaption.
\end{itemize}

The remainder of the paper is arranged as follows. In Section II, we introduce the relevant systems in accessible telerobotics, telepresence, and feedback-based navigation. Section III provides an in-depth explanation of our proposed system. Section IV describes the interaction events during navigation in the simulated warehouse. Section V outlines the limitations of our proposed model and future initiatives. Finally, Section VI concludes the paper by combining key insights and contributions.

\section{Related Research} 
\subsection{Navigation Aids for VI Users}
Individuals with vision impairments have traditionally employed traditional mobility aids like white canes and guide dogs.  Although these instruments can detect and avoid obstacles, they only provide limited information about the surroundings \cite{kuriakose2020multimodal}.  Over the last decade, researchers have focused on improving navigation for VI users by developing technology devices that use audio, haptic, and even visual feedback to convey spatial information.  Each modality has its advantages and disadvantages.

\subsubsection{Audio-Based Navigation} 
Auditory signals are especially useful for conveying directional information and environmental context. Loomis et al. discovered that spatialized 3D audio enhances navigation accuracy in comparison to basic spoken commands \cite{loomis1998navigation}. The Personal Guidance System employed 3D acoustic beacons to indicate destination locations, which users preferred over turn-by-turn directions \cite{loomis2005personal}. Audio effectively communicates context-specific information, such as object identity and relative distance \cite{bainbridge2004berkshire, kurosu2013human}. Long-term use of auditory interfaces can be cognitively demanding, and headphones can suppress important ambient sounds for situational awareness \cite{caraiman2017computer}. Bone-conduction headphones can preserve ambient sounds \cite{Katz_2024}, although regulating audio density and timing remains crucial.   Delays or overlapping communications can confuse users and obscure crucial environmental cues.  While audio feedback is effective, it must be used wisely to avoid cognitive overload \cite{kuriakose2020multimodal}.

\subsubsection{Haptic Navigation}

Haptic feedback is a discreet, non-intrusive navigation aid that does not disrupt auditory perception \cite{Ricci2023Navigation}.   Vibrating belts and wristbands can provide orientation indications based on their location on the body \cite{bourbakis2008multimodal, Kärcher2012sensory}.  For example, vibrations on the left side may indicate an obstacle or a turn to the left.   Haptic devices have limited ability to provide specific spatial data due to poor bandwidth and their dependence on the user's tactile sensitivity \cite{Parshotam2022Exploration}. Attempts to improve this include Braille displays or multi-pin tactile arrays that convey spatial layouts or item outlines \cite{kuriakose2020multimodal, kim2019towards}.  A hybrid system that combines haptic alerts and a tactile visual display to deliver detailed information about nearby impediments. However, these systems need training and time to become familiar with, making them less ideal for real-time decision-making than auditory cues \cite{kuriakose2020multimodal}.

\subsubsection{Visual and Multimodal Aids}

Some VI individuals possess residual vision, urging research into visual augmentations such as high-contrast displays or AR overlays \cite{Gopalakrishnan2020Use, Al-Ataby2016Visual}. These serve as supplemental aids rather than standalone tools. Furthermore, systems frequently convert visual information(e.g., object recognition) into aural or haptic feedback, bridging sensory modalities. Multimodal systems, like EyeBeacons, combine auditory, tactile, and visual cues to provide flexible and redundant navigation support \cite{van2019communicating}.  This redundancy increases safety and user adaptability in a variety of contexts.  However, efficient multimodal integration requires precise coordination to avoid sensory interference.  In user trials, simultaneous audio and haptic cues without noticeable separation created confusion \cite{van2019communicating}.  Recent studies have emphasized that many proposed systems are still in prototype stage and have not undergone substantial real-world testing with varied VI populations \cite{kuriakose2020multimodal}. There is still a need for powerful, user-specific multimodal technologies that solve daily navigational tasks fully.

A prominent recent example is Stanford's Augmented Cane, which uses LiDAR, GPS, and computer vision to provide multimodal feedback via haptic cues on the cane's handle.  This approach goes beyond simple barrier recognition to include semantic interpretation and path planning.  It demonstrates how robotic sensing may improve traditional aids with greater spatial awareness while maintaining a lightweight and user-friendly design \cite{slade2021multimodal}.

\begin{figure*}[t]
  \centering

  \begin{subfigure}[t]{\textwidth}
    \centering
    \includegraphics[width=\linewidth, height=7.5cm]{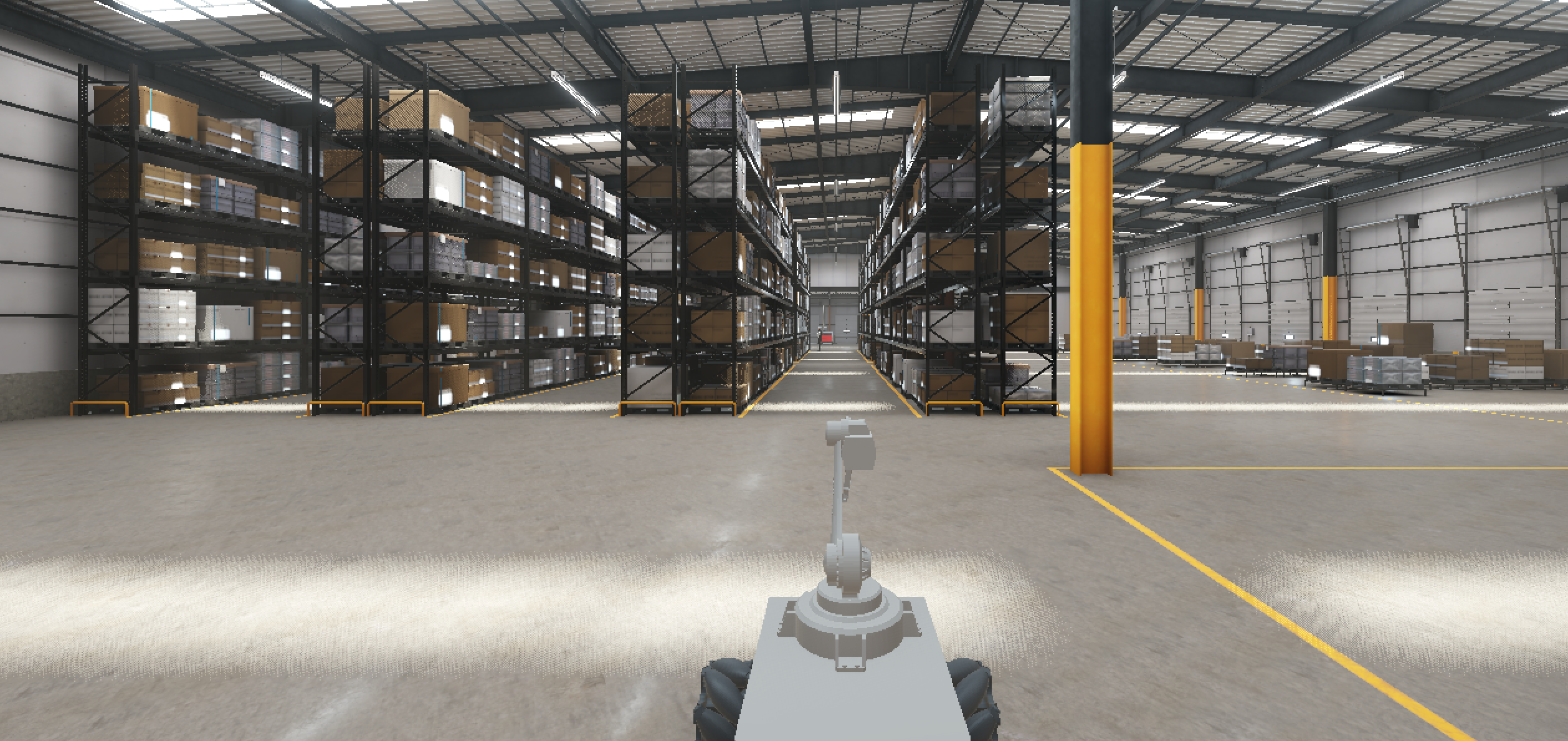}
    \caption{}
    \label{fig:top_view}
  \end{subfigure}
  \vspace{0cm}

  \begin{subfigure}[t]{0.3\textwidth}
    \centering
    \includegraphics[width=\linewidth,height=5cm]{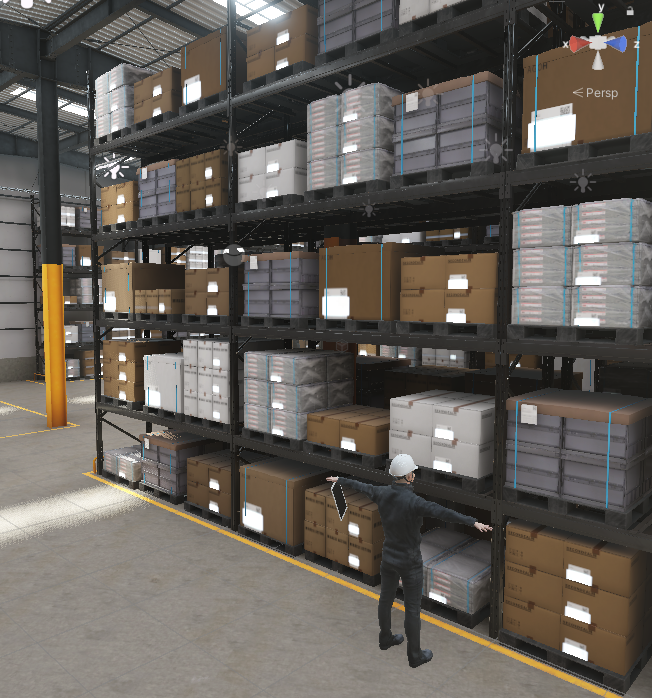}
    \caption{}
    \label{fig:worker}
  \end{subfigure}
  \hfill
  \begin{subfigure}[t]{0.3\textwidth}
    \centering
    \includegraphics[width=\linewidth, height=5cm]{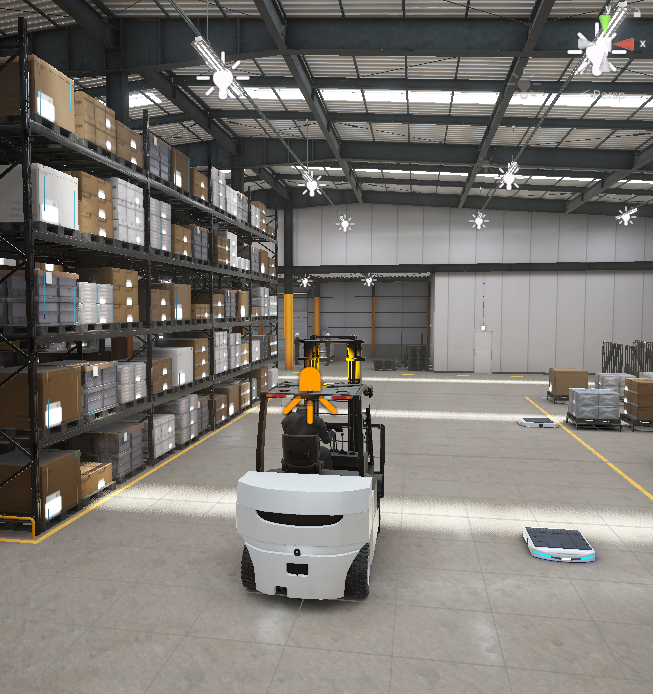}
    \caption{}
    \label{fig:robot_forklift}
  \end{subfigure}
  \hfill
  \begin{subfigure}[t]{0.3\textwidth}
    \centering
    \includegraphics[width=\linewidth, height=5cm]{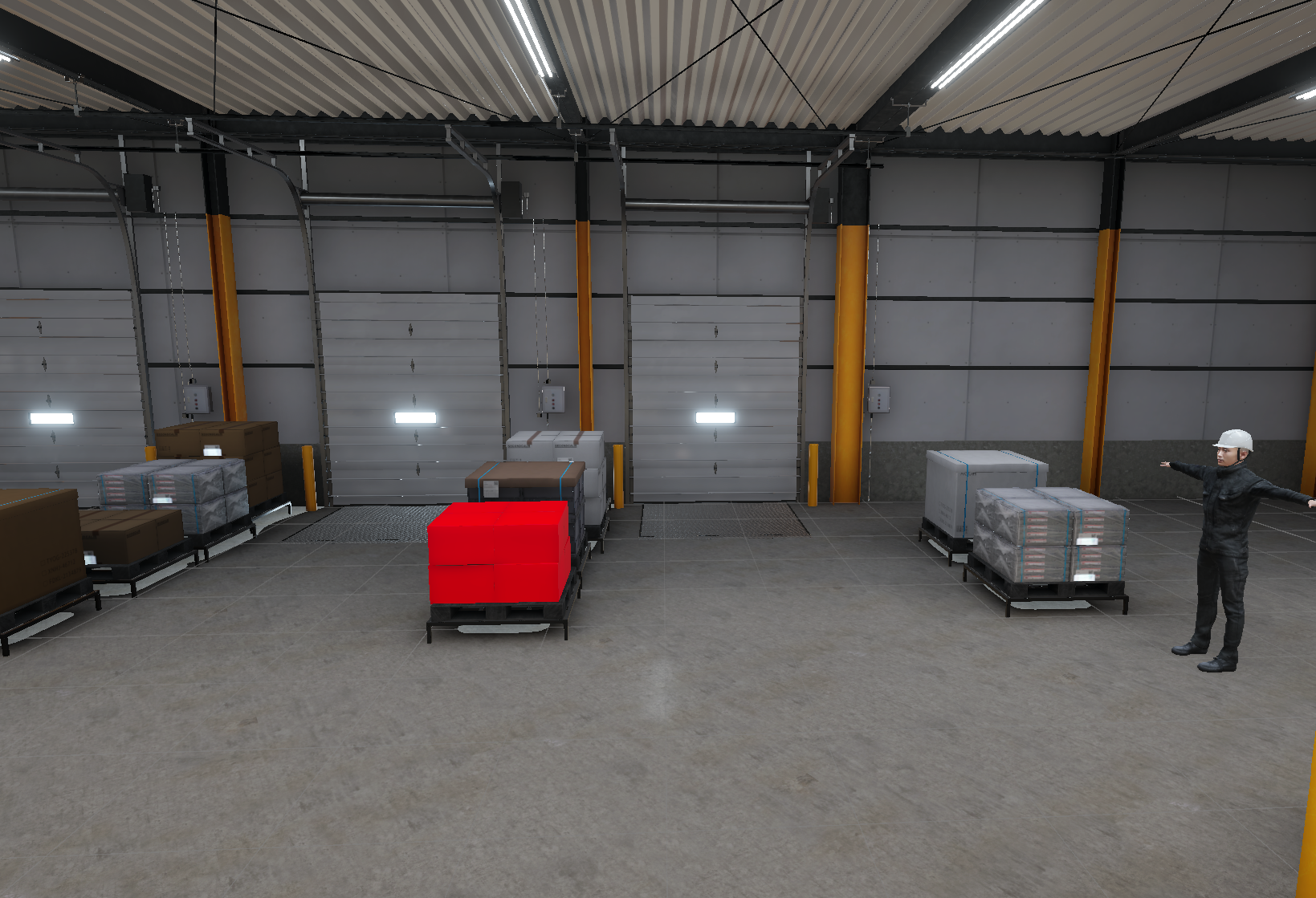}
    \caption{}
    \label{fig:loading_dock}
  \end{subfigure}

  \caption{Layout of the simulated warehouse (a) Robot’s forward warehouse view, (b) Shelves with worker, (c) Forklift and pallet robot (d) Goal represented as the red box. }
  \label{fig:warehouse_views}
\end{figure*}

\subsection{Accessible Telerobotics and Telepresence for VI Users}

Telerobotic devices have the potential to increase VI users' access to remote locations. Currently, most telepresence systems presume the user can understand video feeds, which limits usage for blind individuals \cite{zhang2022telepresence}. Recent studies have attempted to overcome this hurdle by establishing non-visual feedback systems.

Park and Howard's studies were among the first to show how a blind person may "feel" a remote world using a telerobotic interface. In a 2012 study, they proposed a framework in which a mobile robot equipped with depth sensors surveys the surroundings, and a force-feedback haptic device (controlled by the VI user) presents the 3D spatial information in real time \cite{park2012real}. The user holds a stylus or knob that the system moves with resistive forces matching to remote barriers and surfaces, allowing them to explore a room by touch from afar \cite{park2012real}. In subsequent work, they improved this into a haptic visualization system that translates depth camera data (e.g., from a Microsoft Kinect)  into a tangible 3D experience. Experiments demonstrated that blindfolded and blind participants could create a mental map of a remote room's architecture by moving the haptic stylus and feeling where virtual "walls" and objects were situated \cite{park2015telerobotic, park2014haptic}. Kim et al. investigated transforming camera data into dynamic Braille-like displays \cite{kim2019towards}. While these systems provide a high level of spatial detail, they are frequently limited by update speed and device resolution. To solve these limits, several systems integrate audio and haptic feedback, with the former providing contextual information and the latter providing spatial awareness.

Maintaining situational awareness without overwhelming the user is a major challenge in BLV teleoperation. Unlike sighted users, who easily comprehend a visual feed, blind users must synthesize information sequentially. Shared autonomy, in which the robot helps with navigation or collision avoidance while the user makes decisions, has emerged as a viable option. Kamikubo et al. found that VI users prefer keeping control ("boss mode") over helpful robot interventions ("monitor mode") for safety and social navigation \cite{kamikubo2025beyond}. This balance empowers people while increasing safety and efficiency.

Despite advances in assisted navigation and accessible telepresence, few studies have focused on telerobotics for visually impaired individuals in industrial environments such as warehouses.  Previous research has primarily focused on urban mobility, household surroundings, and social telepresence.  Warehouses present particular issues, such as dynamic impediments (e.g., forklifts), complex layouts, and the necessity for task execution (e.g., picking or scanning items).  Traditional assistance, such as canes and smart wearables, are insufficient in such large-scale, fast-paced environments.

Assistive robots have been studied for semi-structured situations such as grocery shops \cite{gharpure2008robot}, although they typically use autonomous navigation rather than user-controlled teleoperation.  To our knowledge, no existing solutions allow VI people to actively steer a robot in a warehouse using multimodal feedback.  Recent industrial attempts show practical interest, but users cannot directly operate these robots \cite{Gomes}.  Research has yet to produce an entire framework that gives a VI user the freedom to explore, navigate, and accomplish tasks in such worlds.

\section{System Overview}
This section addresses the comprehensive system architecture. We discuss the warehouse layout, the modeling of objects and obstacles with Unity's integrated system, the inclusion of navigational paths through NavMesh, and the obstacle avoidance mechanism of along the route. In addition, we analyze the feedback mechanism that encompasses haptic and auditory feedback.
    \subsection{Simulated Environment Design}
        \subsubsection{Warehouse Layout} 
        \begin{table*}[ht]
\caption{Components of the Simulated Warehouse Environment (Static and Dynamic Elements).}
\begin{tabularx}{\textwidth}{p{3cm}p{3cm}X}
\hline
\textbf{Component Category}       & \textbf{Component} & \textbf{Properties}                                                                                  \\ \hline
\multirow{5}{*}{\textbf{Static}}  & Shelves            & Defined as non-walkable NavMesh obstacles and triggers audio feedback.                               \\
                                  & Walls              & Serve as boundary of the layout.                                                                     \\
                                  & Boxes              & Defined as static obstacles, and equipped with box colliders for dynamic interactions.               \\
                                  & Floor              & Flat navigable surface and marked as walkable in the NavMesh. No physical collider has been applied. \\
                                  & Pedestals          & Supporting structures that work as bases for the boxes.                                              \\ \hline
\multirow{3}{*}{\textbf{Dynamic}} & Workers            & Simulated entities that work as static obstacles and trigger proximity-based feedback.               \\
                                  & Pallet Robots      & Simulated mobile units with rigidbody components and motion scripting that act as moving obstacles.  \\
                                  & Forklifts          & Dynamic vehicle entities that follow scripted navigation routes and also act as moving obstacles.    \\ \hline
\end{tabularx}
\label{tab:components}
\end{table*}

To develop the simulated warehouse, we have used a logistic warehouse as the base platform from the Unity Asset Store \cite{unityWarehouse2024}. The project generates high fidelity and cutting-edge graphics by means of High Definition Render Pipeline (HDRP). It is incompatible with the Universal Render Pipeline (URP) and calls Unity version 2022.3.16f1 or above. As shown in the Fig \ref{fig:warehouse_views}, the basic model consists of components of a logistic warehouse including shelves, forklifts, people, pallet robots, and boxes. Table \ref{tab:components} lists the warehouse elements together with their characteristics.

        \subsubsection{Object and Obstacle Modeling} 
        Unity's native 3D primitives and imported asset prefabs help to model objects in the warehouse simulation. Box colliders enable static entities—such as boxes on shelves, pedestals, and workers—to be static obstacles to participate in collision detection. Furthermore equipped with rigid-body component, dynamic items such as pallet robots and forklifts can operate as dynamic obstacles as shown in Table \ref{tab:components}. In the warehouse simulation, these representations enable proximity-based haptic feedback.
        \subsubsection{Controller Input and Mapping} 
        We utilized Unity's Input System \cite{unityinputsystem} which allows users to control the movement of the robot through DualSense controller. This Input System package enbales one to use any type of Input Device to regulate the Unity content. We utilized "Move" from the pre-configured actions and mapped it with the left joystick  to regulate the robot's mobility. The joystick's vertical (Y) axis regulates its forward and backward motion, and the horizontal (X) axis handles its turns to the left or right. We developed the system to use only one joystick for navigation in order to reduce cognitive burden and maximize accessibility for VI users. This method corresponds to the results of accessible game design \cite{gnecco2023increasing} in which user-simplified input improves accessibility for VI people.

    \subsection{Navigation System} 
        \begin{figure}[t] 
          \centering
          \includegraphics[width=\columnwidth]{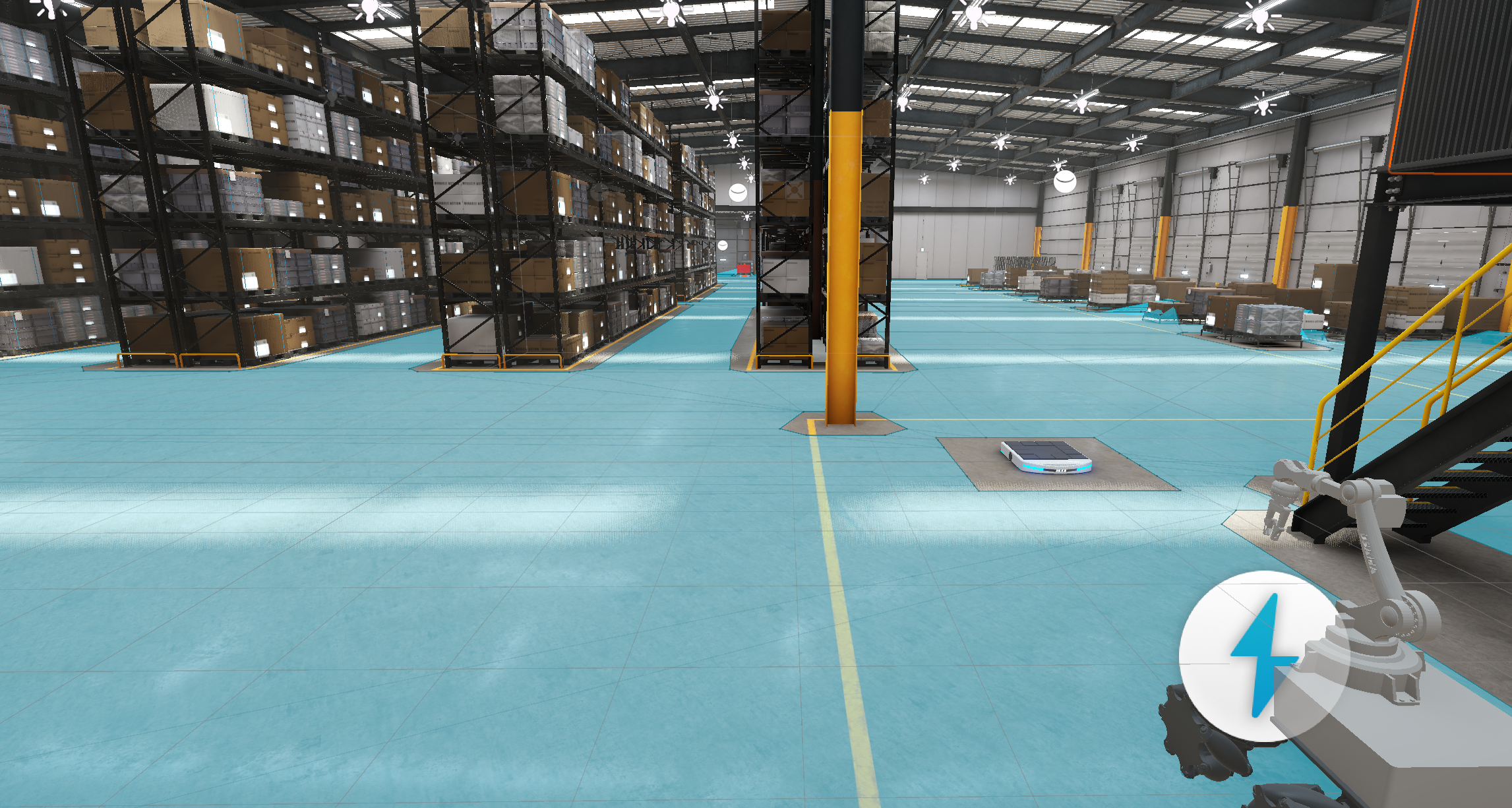}  
          \caption{Generated NavMesh showing the walkable areas (colored in cyan) and obstacles in the warehouse environment.}
          \label{fig:navmesh6}
        \end{figure}
    
        \subsubsection{NavMesh Generation} 
        

We discretize the warehouse interior into a 3‑D voxel height‑field and employ the Recast pipeline to convert that field into a polygonal navigation mesh as shown in Fig. \ref{fig:navmesh6}. Recast first performs a sweep‑and‑prune rasterisation of every triangle in the static scene into 20 cm‑square voxels, then marks walkable cells whose local surface slope is below 45°. To guarantee traversability for the 0.35 m‑radius robot, the height‑field is morphologically eroded by $r = \lceil R_{\text{agent}} / s \rceil$ voxels, where s is the voxel size; the operation is formally identical to a Minkowski sum\cite{das2018computing} and eliminates corridors narrower than twice the agent radius – a safety margin recommended in Recast’s design documentation \cite{recastnav2023}.
Next, a 3‑D Euclidean distance transform fills each walkable voxel with its squared distance to the nearest obstacle. A watershed partitioning of that scalar field produces approximately convex 2‑D regions that are subsequently traced into contours and triangulated via a constrained Delaunay criterion, yielding a planar graph $G =(V, E)$ whose vertices are triangle centroids and whose edges connect adjacent triangles. Dynamic elements are represented at run time by cylindrical obstacle volumes that carve temporary holes in the mesh. As carving does not expand walkable space, we issue an incremental rebuild every $\Delta t = 2 s$ for more than 1\% of the changed carved area, ensuring that the graph remains a valid representation without incurring the $O(|V|)$ cost of a full bake.

        \subsubsection{Pathfinding and Goal Assignment} 
        The problem of transferring the robot from its start configuration $q_s$ to the goal 
$q_g$ reduces to finding a minimum‑cost path, $\pi = <v_0, ...,v_k>$ in $G$. We employ the classical A* search strategy of Hart, Nilsson and Raphael \cite{hart1968formal}  with the cost function,
\begin{equation}
f(v) = g(v) + h(v) = g(v) + \alpha \|p(v) - p(q_g)\|_2
\end{equation}


where $g(v)$ is the accumulated Euclidean cost along $\pi$ and $h$ is an admissible and consistent heuristic scaled by $\alpha=1.$ $A^*$ thus preserves optimality while operating in $O(|E| log |V|)$ time in expectation. Once a sequence of triangles has been recovered, we apply the Simple Stupid Funnel (SSF) algorithm.  The SSF treats the sequence of triangle portals as a corridor and iteratively tightens a left and right funnel around the agent’s line of sight; when the apex of the funnel crosses an edge, a globally shortest poly‑line inside the corridor is produced in $O(k)$ time. Corner points of this line become way‑points $w_i$ and the robot is considered to have progressed to $w_{i+1}$  once its Euclidean distance to $w_i$ falls below 1 meter.
The environment is non‑stationary. Let $P(t)$ denote the cost‑optimal path at time t. Whenever a carved hole invalidates part of  $P(t)$ or the robot’s forward velocity v falls below $0.1\ \mathrm{m\,s^{-1}}$ for more than 1s, we trigger a fresh $A^*$ query, thereby implementing a receding‑horizon policy with period $\tau = 2s$. Because each query expands only $\approx 300$ nodes,the additional computational burden is negligible ($<1 ms$ per cycle) relative to the physics simulation.

        \subsubsection{Obstacle Detection and Avoidance} 

Combining with the pathfinder algorithm described in the previous section, we have come up with an obstacle detection and avoidance mechanism. This mechanism makes sure that the robot can be navigated safely in a warehouse environment with both static and moving obstacles. It keeps recalculating robot's optimal path to destination within a defined timeframe to avoid collisions, enabling a safe traversal towards destination.

We utilized Unity's Navmesh system for modeling the warehouse environment. During the robot's movement, obstacles may move and appear in its path. We recalculate the robot's optimal path with our predefined algorithm every 2 seconds to keep the current valid path and provide right navigation guidance to the user. This dynamic path adjustment ensures that the robot's navigation is consistently adapted to changing environment, avoiding both static shelves and moving pedestrians and forklifts.

Initially, the NavMesh is baked with the warehouse's map and walkable paths are identified. Several static objects such as shelves and other structures are labeled as non traversable. So, when the robot's path is generated, the static obstacles are already avoided and if a use case arises when the operator has moved the robot too close to a shelf, we play an audio cue to notify them. At the same time, the system recalculates updated path towards the goal to prevent the robot from getting stuck and and ensure it moves towards the goal. With this simple and intuitive technique the navigation system handles staic obstacles.
Unity's NavMeshObstacle is used to handle the moving objects in the warehouse. NavMeshObstacle carves temporary gaps into the NavMesh. These change the walkable areas and thus require updated optimal path towards the goal which is generated in real-time within the path recalculation time frame. Each moving object $o \epsilon O(t)$ where $O(t)$ denotes the set of obstacles at time $t$ has a cylindrical volume denoting its position and volume it takes. The NavMesh recalculation is triggered when more than 1\%  of the mesh is changed due to obstacle movement. This ensures that the robot can traverse its path avoiding collision.
The robot's current position $P(t)$ is evaluated against $O(t)$ at any given time $t$ to check if the current waypoints intersects any object and if it does, the pathfinding algorithm is launched again to update the route. Figure \ref{fig:obstacle_avoidance} shows the path curved in order to avoid moving forklift.

The optimal path, $\pi (t)$ starting from the current $p(t)$ position to the destination $g$ is calculated periodically using the $A^*$ algorithm to keep up with the moving obstacles in the system. This process of finding valid path at each timestamp can be denoted as,

\begin{equation}
\pi(t) = \arg \min_{\pi} \sum_{e \in \pi} c(e) \quad \text{where, }   \mathbf{p}(t) \notin O(t)
\end{equation}

where $c(e)$ represents the cost (e.g., Euclidean distance) associated with traversing an edge 
$e \epsilon E$, and the path $\pi(t)$ must avoid any obstacles at time $t$. The pathfinding algorithm ensures that if an obstacle is encountered, the route is recalculated in real-time while minimizing path cost, ultimately ensuring that the robot can still reach its destination safely.
The system looks for potential obstacles at each waypoint as the robot passes through a sequence of those. The path is recalculated if the next waypoint is blocked, guaranteeing that the robot always travels the best path to its destination. As the robot moves closer to the goal, the obstacle detection and avoidance procedure is carried out iteratively, guaranteeing that the robot is dynamically rerouted when required. Algorithm \ref{alg:optimal_pathfinding} sums up the overall process.

\begin{algorithm}[ht]
\caption{Optimal Pathfinding with Obstacle Avoidance}
\label{alg:optimal_pathfinding} 
\begin{algorithmic}[1]
\STATE Initialize the NavMesh with static obstacles (walls, shelves).
\STATE Set the robot's initial position \( \mathbf{p}_0 \) and destination \( \mathbf{g} \).
\STATE Compute the initial optimal path \( \pi_0 \) using the A* algorithm:
\[
\pi_0 = \arg \min_{\pi} \sum_{e \in \pi} c(e) \quad \text{where, }  \mathbf{p}_0 \notin O(0),
\]
where \( c(e) \) represents the cost of traversing edge \( e \) and \( O(0) \) is the set of static obstacles at the beginning
\STATE Divide the path \( \pi_0 \) into a series of waypoints \( w_1, w_2, \dots, w_n \).
\WHILE{robot has not reached the destination \( \mathbf{g} \)}
    \STATE Move the robot towards the next waypoint \( w_i \).
    \STATE Monitor the environment for new obstacles using Unity's physics engine.
    \IF{a new obstacle is detected at \( \mathbf{p}_o \) such that \( \|\mathbf{p}_o - \mathbf{p}(t)\| < R_{\text{threshold}} \)}
        \STATE Recalculate the path \( \pi(t) \) using:
        \[
        \pi(t) = \arg \min_{\pi} \sum_{e \in \pi} c(e) \quad \text{where, }  \mathbf{p}(t) \notin O(t),
        \]
        where \( O(t) \) represents the dynamic obstacles at time \( t \).
        \STATE Update the path with new waypoints.
        \STATE Recheck for obstacles along the updated path.
    \ENDIF
    \IF{the robot reaches waypoint \( w_i \)}
        \STATE Update to the next waypoint \( w_{i+1} \).
    \ENDIF
\ENDWHILE
\IF{the robot's position \( \mathbf{p}(t) \) reaches the destination \( \mathbf{g} \)}
    \STATE Terminate the navigation.
\ENDIF
\STATE If a new obstacle blocks the path during movement, repeat steps 3 to 6 to adjust the route.
\end{algorithmic}
\end{algorithm}

    \subsection{Feedback Mechanism} 
        \subsubsection{Haptic Feedback} 
\begin{figure*}[ht]
  \centering

  \begin{subfigure}[t]{0.31\textwidth}
    \centering
    \includegraphics[width=\linewidth]{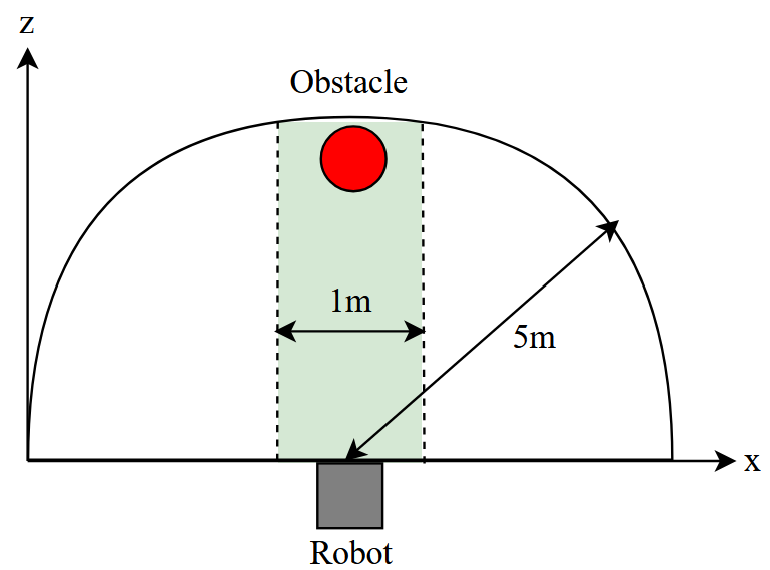}
    \caption{}
    \label{fig:left}
  \end{subfigure}
  \hfill
  \begin{subfigure}[t]{0.31\textwidth}
    \centering
    \includegraphics[width=\linewidth]{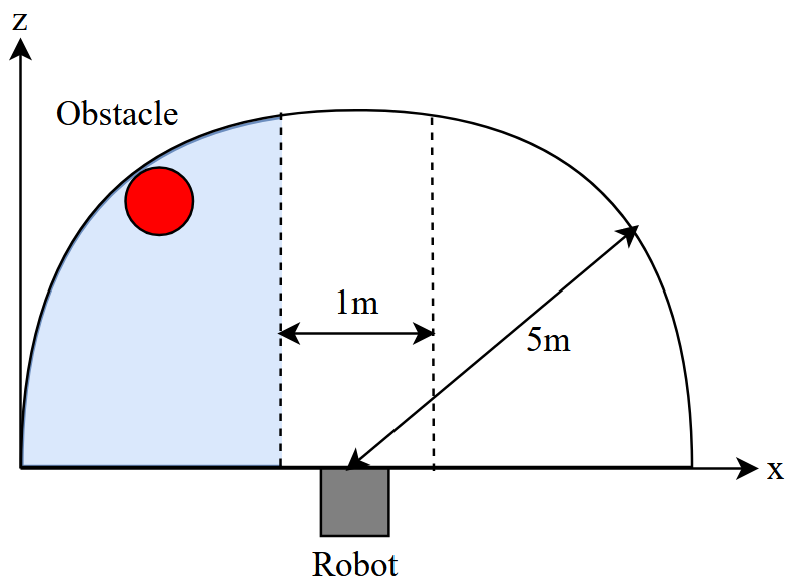}
    \caption{}
    \label{fig:center}
  \end{subfigure}
  \hfill
  \begin{subfigure}[t]{0.31\textwidth}
    \centering
    \includegraphics[width=\linewidth]{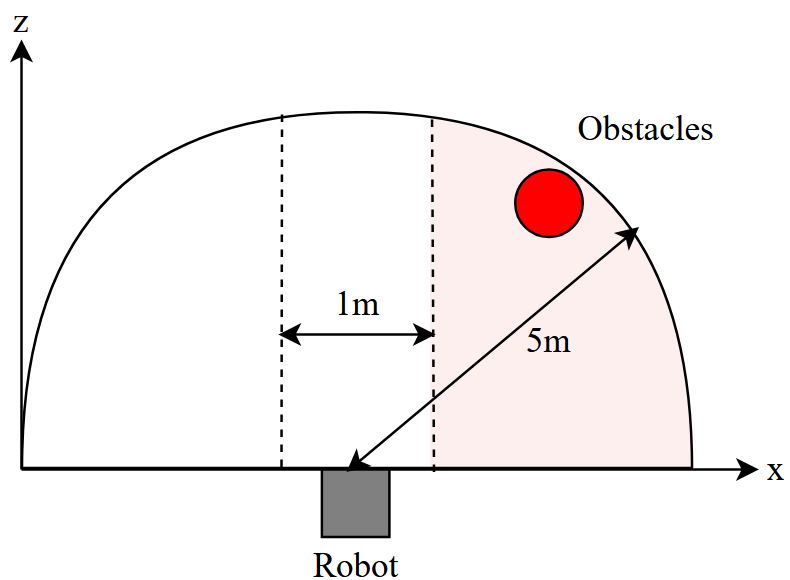}
    \caption{}
    \label{fig:right}
  \end{subfigure}
  \hfill
  \begin{subfigure}[t]{0.9\textwidth}
  \centering
    \includegraphics[width=\linewidth]{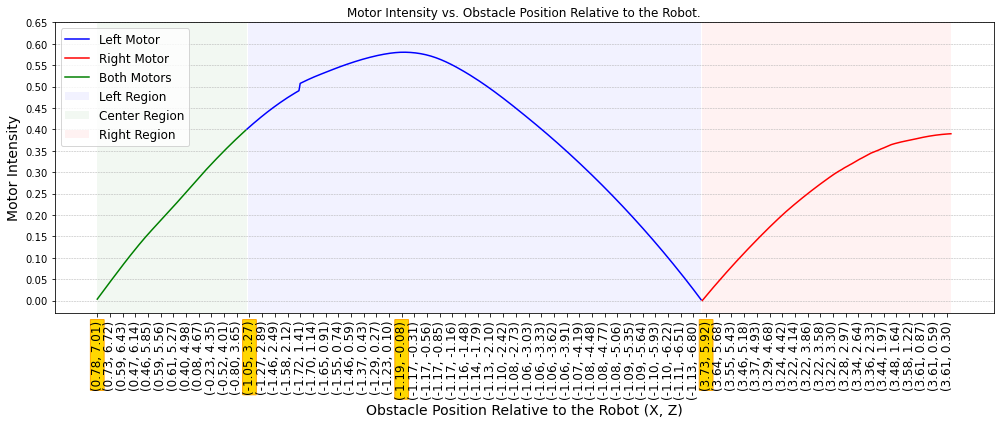}
    \caption{}
    \label{fig:right}
  \end{subfigure}

  \caption{Illustration of haptic feedback zones and motor intensity response based on the obstacle's relative position to the robot. The obstacle is positioned at (a) center, (b) left, and (c) right, relative to the robot. (d) The X-axis represents the obstacle's position within the robot's local frame (X and Z only, as Y is constant), and the Y-axis indicates the motor intensity values. As the obstacle approaches the robot (i.e., distance decreases) or vice versa, the intensity of the relevant motor increases. Initially, the obstacle is in center to the robot and both motor vibrates with equal intensity as indicated by the green curve. As the distance between them diminishes, the intensity accelerates. Then the obstacle is positioned in left to robot and only left motor vibrates. The intensity gradually drops as the obstacle goes beyond the detection perimeter. This is shown as the blue curve. The robot encounters another obstacle on its right side, causing only the right motor to vibrate. As the distance between them decreases, the intensity increases which is illustrated by the red curve.}
  \label{fig:haptic_feedback}
\end{figure*}

        We employed haptic feedback to develop tactile cues to aid proximity-based obstacle detection and avoidance in the warehouse simulation. For VI individuals haptic feedback is an important tool for navigating their immediate surroundings. In this research, we focused on developing proximity based haptic feedback to enable users to perceive the location of obstacles and evade them during navigation.  According to current research, users may find pattern-based haptic cues challenging or unclear in real-time circumstances \cite{kuriakose2020tools}. As a result, we opted for dynamically scaled haptic intensity instead of the patterned vibrations. 
In the simulation, the robot maintains a defined radial detection boundary of 5 meters for which haptic alerts are given. When a static or moving obstacle enters into this detection perimeter, the system starts relaying haptic feedback based on the relative position of the robot and obstacle. The position of the nearest obstacle is converted into the robot’s coordinate space using \texttt{player.InverseTransformPoint}. To determine the position of an obstacle relative to the robot—center, left, or right—we defined a threshold named \texttt{centerRange}, set it to 1 meter, and compared it against the obstacle's x-coordinate value, \texttt{local.Position.x}. If \texttt{localPosition.x < -centerRange}, the obstacle is on the left, and only the left motor vibrates. Whenever \texttt{localPosition.x > centerRange}, the obstacle is located to the right, and only the right motor vibrates. Obstacles within the range \texttt{[-centerRange, +centerRange]} are considered to be in center, causing both motors to trigger with same level of intensity. Fig. \ref{fig:haptic_feedback} illustrates the above approach with a position vs intensity plot. 

To generate haptic intensity levels, we applied a logarithmic function, as the Weber-Fechner rule states that the perceived intensity of a stimulus increases in proportion to the logarithm of its real physical intensity \cite{yokoe2025intuitive}.

\begin{equation}
H(d) = \log\left(1 + (1 - \frac{d}{d_{\text{max}}})\right)
\end{equation}

The function enables the calculation of haptic intensity $H(d)$ based on the current distance $d$ from the robot to the nearest obstacle.  $d_{\text{max}}$ represents the defined radius of the detection perimeter surrounding the robot. Initially, we normalized the distance between the robot and the obstacle over the maximum detection radius. The normalized value is then inverted, and the natural logarithm is applied to it to output a smooth decaying intensity curve as obstacles approach or move away. The result of this formula is a perceptible change of haptic intensity depending on the varying relative distance of the obstacles. There is a noticeable high intensity vibration when obstacles are nearby and a gradual decrease of intensity as it moves away. The inverse is also true for when an obstacle approaches the robot. 

        \subsubsection{Audio Feedback} 

VI people heavily rely on audio feedback in their daily living activities such as reading books\cite{jackson2012audio}, browsing the Internet\cite{roth2000audio, shimomura2010accessibility}, operating smartphones\cite{rodriguez2014accessible, khan2021insight}, crossing roads\cite{jin2020acoussist, mascetti2016sonification, wiener1997use}, and so on. This aspect can be beneficial in the context of telerobotic system when a VI user is navigating a robot in a warehouse. In this study, we integrated audio cues across various settings to facilitate safe navigation for the operators.

We employed a Text-to-Speech (TTS) technology to deliver audio instructions. VI users prefer rapid TTS systems \cite{cheema2024user} as they often require immediate responses based on auditory feedback. We also implement clock positions, also known as clock-based directions or clock bearings, which uses the analogy of a 12-hour clock face to describe the relative direction of an object or location.

To determine the robot's direction and convert it to clock face direction, we first calculate the direction vector of current position, $P_{\text{cur}} = (x1,y1,z1)$ and the destination $P_{\text{dest}} = (x2,y2,z2)$ given by distance vector,
\begin{equation}
d = P_{\text{dest}} - P_{\text{cur}} = (x2-x1, y2-y1, z2-z1)
\end{equation}
This distance vector is then normalized by,
\begin{equation}
    d_{\text{norm}} = d / \lvert d \rvert
\end{equation}
where,
\begin{equation}
\label{eq:eucledian}
    \lvert d \rvert = \sqrt{(x_2-x_1)^2 + (y_2-y_1)^2 + (z_2-z_1)^2 } 
\end{equation} 
Equation \ref{eq:eucledian}  denotes the Eucledian distance \cite{hughes2013computer} between $P_{\text{cur}}$ and $P_{\text{dest}}$.
To transform the destination vector $d$ into  robot's local coordinate system we used Unity's InverseTransformDirection taking the robot's orientation into account. To compute the angle $\theta$, we use the atan2 function,
\begin{equation}
\theta = atan2(d_x, d_y). \frac{180}{\pi}
\end{equation}
 where $d_x$ and $d_y$ are the components of the direction in the robot's local x and z axes.
Then, $\theta$ is normalized to the range $[0^{\circ}, 180^{\circ}]$ using
\begin{equation}
\theta_{\text{normalized}} = (\theta + 360) \quad mod \quad 360
\end{equation}
Lastly, we divide the normalized angle by $30^{\circ}$ (a clock face has 12 hours, each of which represents $30^{\circ}$) and round the result to the closest integer in order to convert the angle into a clock direction. 

\begin{equation}
    direction_{{\text{clock}}} = \lfloor \theta_{\text{normalized}} / 30 \rfloor
\end{equation}
If $direction_{{\text{clock}}} = 0 $ it is mapped to 12'o clock. 
This approach makes it easier for visually impaired users to understand the robot's movement by explaining it in terms of clockwise direction, such as "3 o'clock," "6 o'clock," etc.
Moreover, in order to reduce collisions with warehouse shelves, we implemented audio cue that is provided when the robot is in close proximity to them. An auditory announcement, "You have reached your destination," is delivered when the robot reaches the goal object. In addition, we establish a threshold to ascertain if the robot is obstructed by an object. If it is inactive over the specified threshold duration, a stuck audio signal is triggered to denote this condition.

        \subsection{Visual Feedback} 
\begin{figure}[htbp] 
  \centering

  \begin{subfigure}[t]{\columnwidth}
    \centering
    \includegraphics[width=\linewidth]{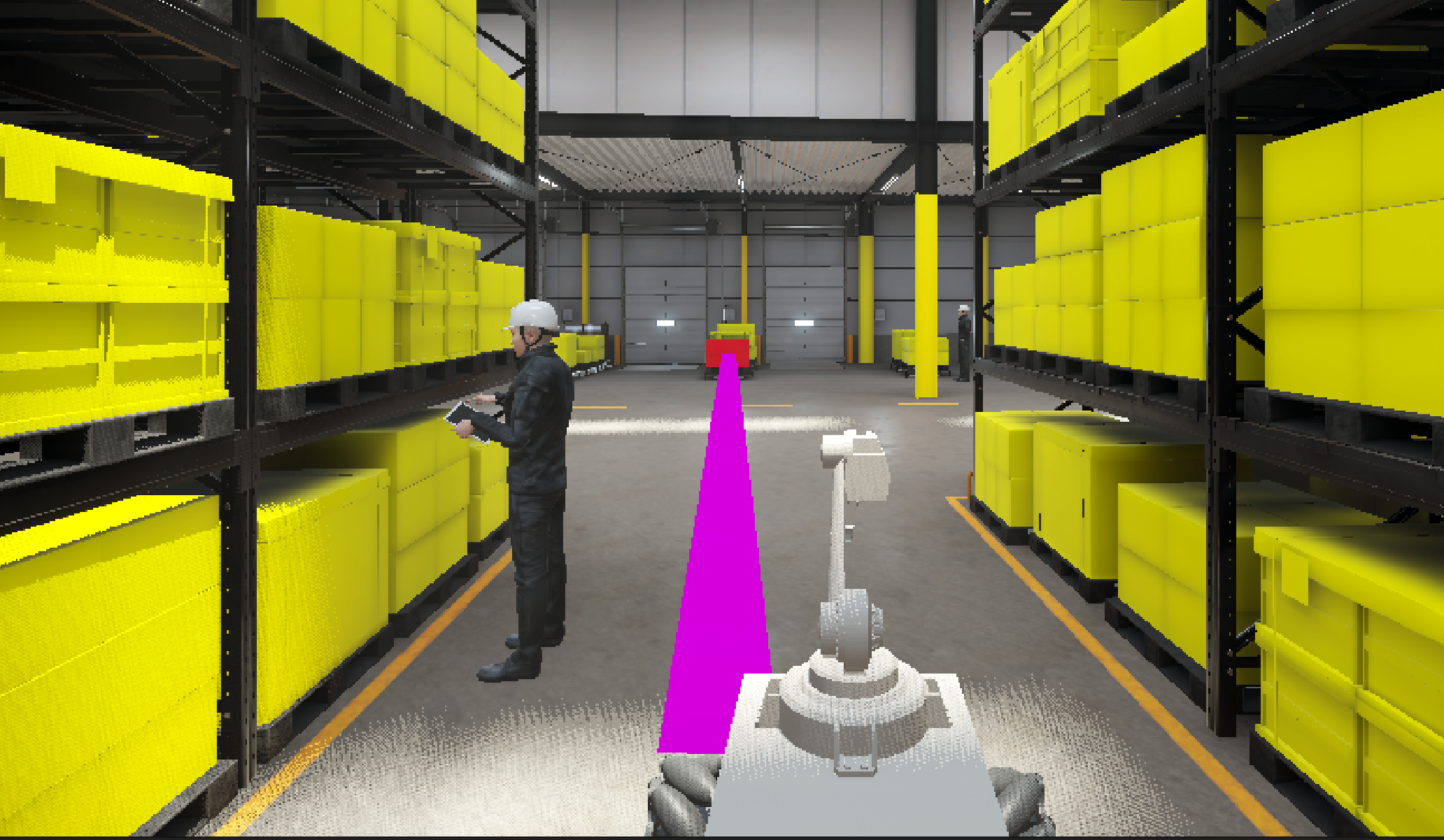}
    \caption{}
    \label{fig:goal}
  \end{subfigure}
  
  \vspace{0.2cm}  
  
  \begin{subfigure}[t]{\columnwidth}
    \centering
    \includegraphics[width=\linewidth]{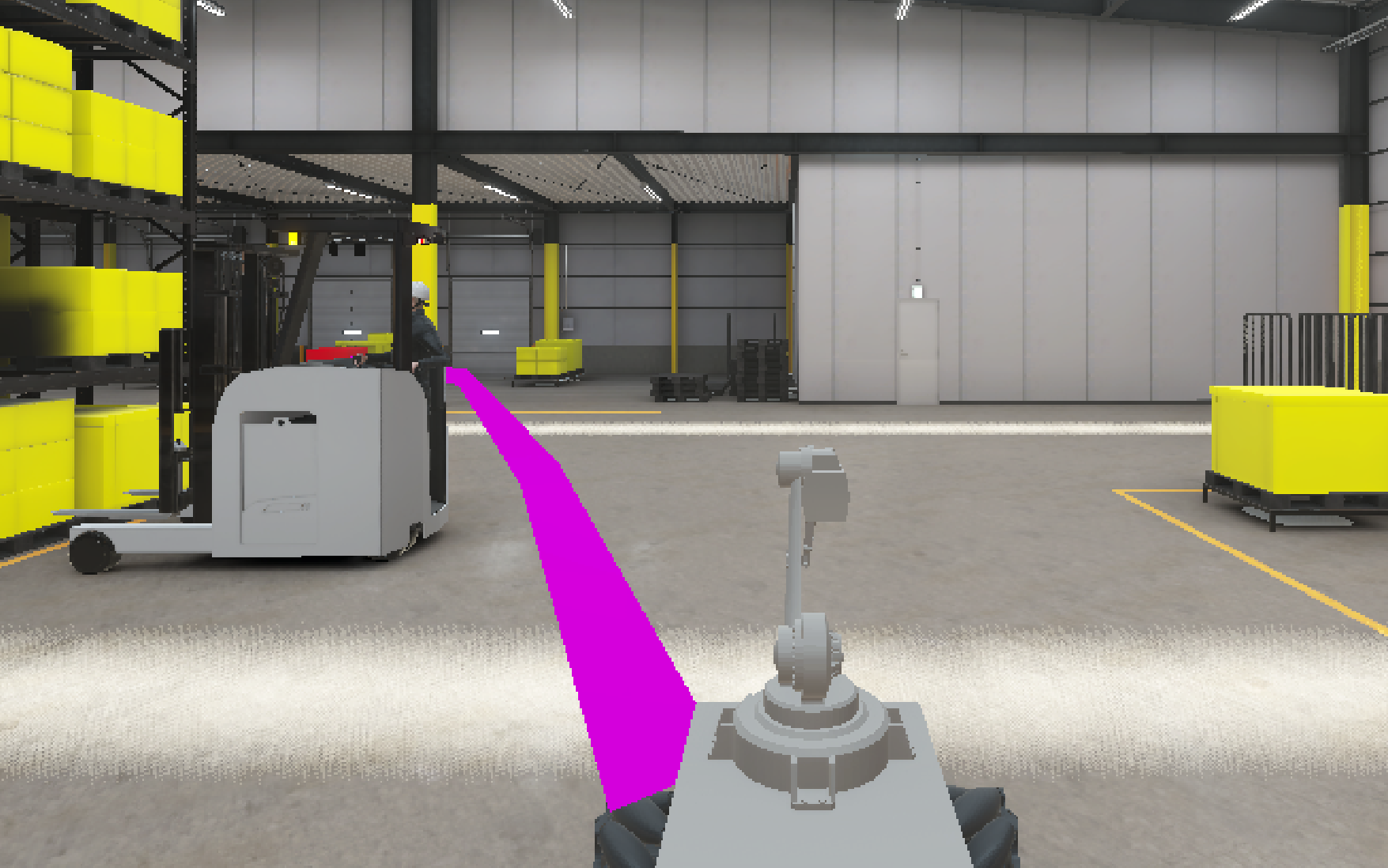}
    \caption{}
    \label{fig:obstacle_avoidance}
  \end{subfigure}
  
  \caption{Illustration of the navigation system with high contrast colors highlighting the boxes, destination object, and navigation path. (a) The robot follows a straight path to the goal guided by the purple visual path. (b) When a dynamic obstacle (forklift in this case) obstructs the route to the goal, our algorithm automatically replans its path to avoid collision, maintaining safe navigation toward the target.}
  \label{fig:navigation}
\end{figure}

To facilitate the low vision operators, we generated a brightly rendered path-line showing the path towards the destination. As described in Algorithm \ref{alg:optimal_pathfinding}
the waypoints and thus the path update dynamically with the changing positions of the moving obstacles to ensure a valid traversal avoiding obstacles. So our rendered path line updates to the latest optimal path making a consistent visual guideline to the users with low vision as shown in Figure \ref{fig:navigation}.
In addition, the boxes on the shelves and the goal object were marked with bright colors to be easily recognized by the operators.
        
\section{Interaction Events and Task Completion}
In this section, we discuss how collision is detected in Unity while navigating through the warehouse, how goal is detected, and how the system handles session completion. Additionally, we discuss the logging process that records total no of collisions and total elapsed time for each user. 
    \subsection{Collision Detection} 
\begin{table*}[t]
  \centering
  \caption{Summary of Interaction Events, Feedback Responses, and Logging Mechanisms in the Simulated Warehouse}
  \label{tab:summary}
  \begin{tabularx}{\textwidth}{p{3cm}Xp{3cm}X}
    \toprule	
    \textbf{Event Type} & \textbf{Trigger Condition} & \textbf{System Feedback} & \textbf{Logging Information}\\		
    \midrule
    Proximity to obstacle & Robot is within 5.0 meter of the obstacle & Haptic based on proximity & Robot position, obstacle position (left, right, center)\\
    Proximity to shelf & Robot is within 1.0 meter of the shelf & Audio & Robot position\\
    Collision with obstacle & Robot contacts with static or moving obstacles & Haptic, Stuck audio & Robot position, obstacle collision count\\
    Collision with shelf & Robot contacts with shelf & Stuck audio & Robot position, shelf collision count\\
    Goal reached & Robot touches the goal & Audio & Total elapsed time to reach goal from start\\
  \bottomrule
\end{tabularx}
\end{table*}

Collision detection in our telerobotic simulation is accomplished with Unity's physics engine - \texttt{OnCollisionEnter} and \texttt{Physics.overlapSphere}. Unity's built-in collision handler, \texttt{OnCollisionEnter} is triggered automatically when a collider associated to one game object comes in contact with a collider of another game object. In our system, Unity passes a collision object when the robot physically runs across another object in the scene. Our method identifies two distinct kind of collisions - collision with shelves and obstacles. It identifies the type by utilizing the \textit{layer} of the relevant item assigned in Unity. If the object belongs to the shelf layer, audio feedback is provided as discussed above, and shelf collision count is recorded in the CSV file for future use. Additionally, if the object is associated with obstacle layer, haptic feedback is provided as described before, and obstacle collision count is recorded in the same CSV file. In addition, we have used \texttt{Physics.OverlapSphere} which is a Unity method to detect nearby collider within a predefined area. Unlike \texttt{OnCollisionEnter} which can only detect objects upon collision, \texttt{Physics.OverlapSphere} can identify objects within a specified radius. It enables proximity-based haptic feedback in the system.

    \subsection{Goal Detection and Session Completion} 
The system keeps checking the relative distance of the robot with the target position to find out if it has crossed the proximity threshold. This threshold is set to one meter from the robot to the destination. Once that threshold is crossed, the system determines it as goal completion and triggers a set of events. At first, an audio announcement is made to notify the user that the goal has been reached after which the vivid line showing the optimal path is cleared from the screen. 
Then the system saves the collision counts i.e. collision with shelves and obstacles as described previously, and total elapsed time from start to goal completion. This information will assist us in evaluating the system through future user studies. Table \ref{tab:summary} provides a summary of events that the robot may encounter during navigation, corresponding feedback responses, and logged information.

\begin{figure}[t] 
          \centering
          \includegraphics[width=\columnwidth]{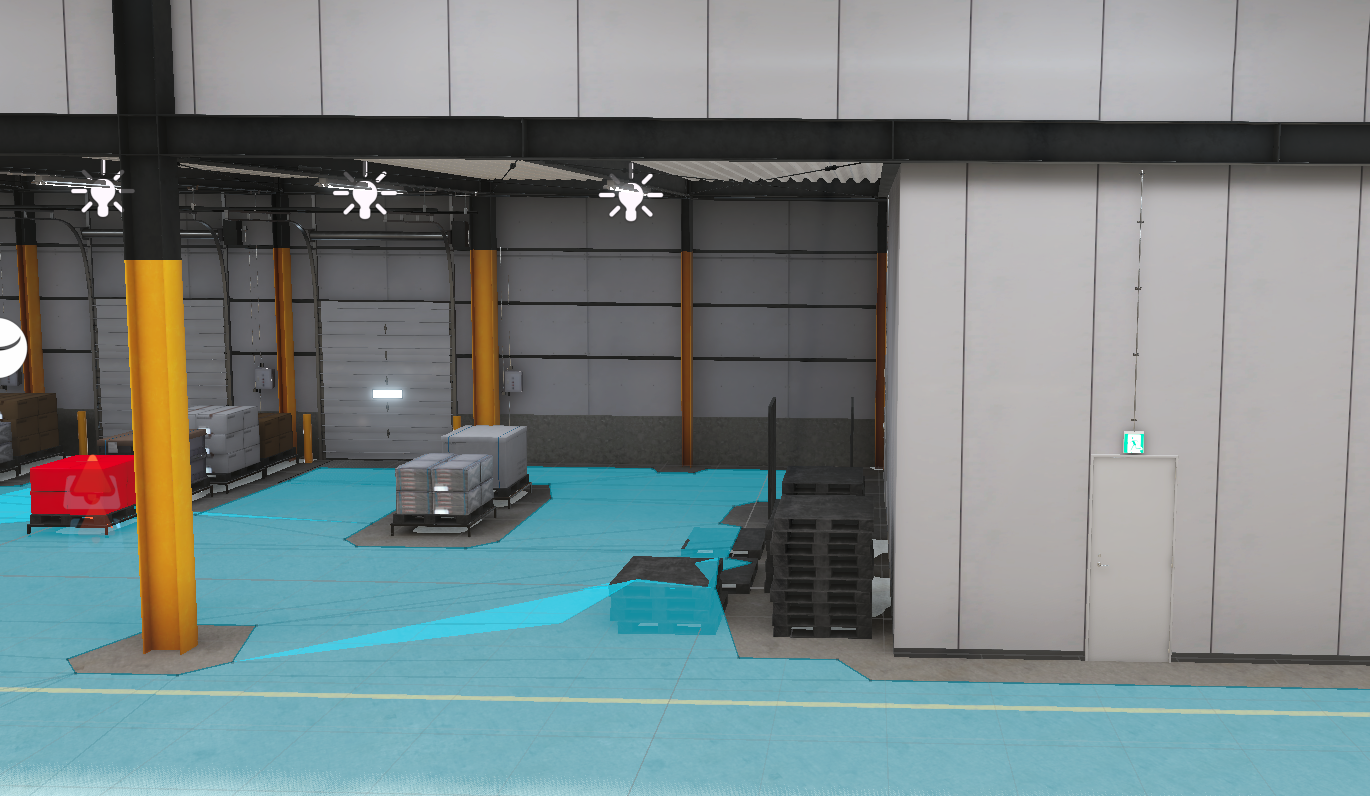}  
          \caption{NavMesh struggles to detect low height obstacle i.e. pedestal.}
          \label{fig:navmesh8}
        \end{figure}

\section{Limitations and Future Work} 
While NavVI has the potential to work as a testbed for accessible telerobotic operation, it possesses several limitations that need to be addressed in future versions. The existing system employs vibration scaling across the three zones outlined in the system overview, with intensity regulated for both motors according to the obstacle's position relative to the robot. Although the haptic intensity curve changes gradually with distance, the DualSense controller initially delivers a sudden vibration before it gradually adjusts to reflect the actual distance of the obstacle. In addition, though our NavMesh system provides reliable path planning with obstacle avoidance, it can struggle with detecting low-height obstacles such as pedestals as shown in Fig \ref{fig:navmesh8}. 

Future work will focus on mitigating these issues to establish a safe navigation system for VI users. Furthermore, we will perform a structured user study with BLV participants to assess feasibility and usability of our approach. We intend to perform comprehensive surveys that will assess the system's utility and any cognitive overload experienced by users to investigate the associated human factors.
\section{Conclusion} 
In this work, we proposed NavVI, a simulation-based telerobotic system intended for facilitating warehouse navigation for VI users through synchronized multimodal feedback. To maintain user control over navigation, the system combines haptic and auditory feedback, and visual cues to improve spatial awareness and assist to avoid obstacles. The simulation provides a robust testbed for building, assessing, and improving inclusive robotic interfaces. For VI users, our system includes proximity-aware haptic feedback depending on obstacle's location relative to the robot, auditory feedback, clock-based direction cues, and additional high-contrast visual indicators specifically for low vision users. To enable future performance analysis and user studies, our system also records interaction events including total no of collisions, and total elapsed time. Future work includes conducting user studies with BLV participants to explore the feasibility of our proposed simulated warehouse environment.

\section*{Acknowledgment} 

This research was partially supported by National Science Foundation (NSF) grant 2438700. The authors express gratitude to Dr. Nicholas Gans, principal research scientist and head of the Autonomation and Intelligent Systems Division of the University of Texas at Arlington Research Institute. He also serves as an Associate Professor in the Department of Computer Science and Engineering at the University of Texas at Arlington. We also convey our gratitude to Raelene Gomes, Senior Manager of IT Operations at Austin Lighthouse (ALH), for assisting us in understanding the surroundings and observing the employees at ALH. In addition, the authors used ChatGPT and Quillbot to structure and paraphrase the sentences in Related Research, Audio Feedback, and Limitations and Future Works sections.
The information presented here originates from the authors' work. The opinions, conclusions, and recommendations presented in this paper are of the author(s) and do not necessarily represent the views of NSF.

\bibliographystyle{IEEEtran}

\end{document}